\documentclass[prl,twocolumn,aps]{revtex4}
\usepackage{graphics}

\begin{document}

\title{Dynamic Model of Super-Arrhenius Relaxation Rates in Glassy Materials}
\author{J. S. Langer}

\author{Ana\"el Lema\^{\i}tre}

\affiliation{Department of Physics,
University of California,
Santa Barbara, CA  93106-9530  USA}

\date{January, 2005}

\begin{abstract}
Super-Arrhenius relaxation rates in glassy materials can be associated with thermally activated rearrangements of increasing numbers of molecules at decreasing temperatures.  We explore a model of such a mechanism in which string-like fluctuations in the neighborhood of shear transformation zones provide routes along which rearrangements can propagate, and the entropy associated with critically long strings allows the rearrangement  to be distributed stably in the surrounding material.  We further postulate that, at low enough temperatures, these fluctuations are localized on the interfaces between frustration-limited domains, and in this way obtain a modified Vogel-Fulcher formula for the relaxation rate.
\end{abstract}

\maketitle

Recent developments in the shear-transformation-zone (STZ) theory of amorphous plasticity \cite{FL,LP}, especially its success in accounting for the behavior of bulk metallic glasses \cite{FLP,Teff}, have prompted us to take a fresh look at the super-Arrhenius rates that characterize relaxation mechanisms in such materials.  In our STZ calculations so far, we have simply deduced these transition rates from measured linear viscosities and used them in predicting, for example, nonlinear plastic responses to driving forces.  To make further progress, it will be useful to have a deeper understanding of the physical mechanism that underlies these rates.  (See \cite{JAMMING} for a recent summary of a wide variety of research in this and related fields.) 

The main assumption of the STZ theory is that a deformable glassy material is fundamentally an elastic solid in which irreversible molecular rearrangements occur at special sites -- so called ``flow defects'' or ``STZ's.''\cite{TURNBULL,SPAEPEN77,ARGON79}  The STZ theory therefore differs from mode-coupling theories \cite{GOTZE91,GOTZE92} whose starting point is a liquid-like model. Our version of STZ theory differs also from the earlier flow-defect theories in that it models the defects as two-state systems that carry information about the internal state of the material.  In this way, the theory describes both the onset of plastic flow at a yield stress and a variety of memory effects. 

In our most recent version of the STZ theory \cite{Teff}, we characterized the configurational degrees of freedom, i.e. the inherent states of the system \cite{STILLINGER}, by an effective disorder temperature which, under nonequilibrium conditions, can be different from the temperature of the thermal bath. We supposed that the STZ's are especially deformable local density fluctuations that are far out in the wings of the effective thermal distribution.  Our implicit assumption was that only a narrow range of these configurational fluctuations is  sufficiently deformable yet populous enough to participate in plastic deformation.  We are concerned here with several different but closely related transformation rates pertaining to the STZ's. Specifically, we want to compute the rate at which STZ's switch from one orientation to another during shear deformations, the rates at which they are created and annihilated, and the rate at which the effective temperature relaxes toward the temperature of the heat bath. More generally, we are interested in the mechanisms by which an amorphous solid makes transitions between its inherent states. 

As has long been recognized in theories of glass dynamics (e.g. see Adam and Gibbs \cite {ADAM-GIBBS}), a qualitative reason why transition rates become anomalously slow at low temperatures is that, as the temperature decreases, the statistically most probable transition states are those that involve increasingly large numbers of molecules.  There have been several recent attempts to make this idea more precise. Notably, Xia and Wolynes \cite{WOLYNES} have postulated that the transitions are nucleated by liquid-like droplets characteristic of an incipient random first-order transition between solid-like and liquid-like (jammed and unjammed) phases.  This postulate, plus some scaling analysis, produces the Vogel-Fulcher result.  An alternative picture has been proposed by Garrahan and Chandler \cite{CHANDLER}, whose transitions are enabled by fluctuating patterns of mobile defects in an otherwise jammed system.  The mobility of these defects -- as opposed to our immobile STZ's -- produces distinctive behavior in this model. In particular, Garrahan and Chandler conclude that relaxation times should obey the Arrhenius formula only at low temperatures rather than, as is more commonly supposed, changing from Arrhenius to super-Arrhenius as the temperature decreases.  So far as we can tell, neither of these approaches attributes enough structure to the transition mechanism to describe the stress-driven change from thermally assisted viscous creep to superplastic flow that emerges from the present STZ theory.  Our purpose here is to examine a transition mechanism that relates directly to our dynamical model of the STZ's.

The model that we shall explore is motivated in part by work of Glotzer and colleagues \cite{GLOTZER1,GLOTZER2}, who discovered in molecular-dynamics simulations that transitions between inherent states in glass-forming liquids take place via motions of string-like groups of molecules. We postulate that, at temperatures low enough that most of the system is tightly jammed, localized molecular rearrangements might be entropically enabled by strings of small molecular displacements that distribute the disturbance throughout larger parts of the material.  In granular materials, our hypothetical mechanism might be visualized as a kinetic fluctuation that allows molecules to undergo small displacements along a force chain.  

For simplicity, consider first just the spontaneous STZ creation rate, that is, the rate at which STZ's are created by thermal fluctuations in the absence of external driving.  It is useful to think of STZ creation as an event in which the glassy analog of a vacancy and an interstitial first form, then move away from one another, and finally stabilize at an indefinitely large separation.  More generally, the formation of an STZ is a spontaneous increase in the configurational disorder of the system, as measured by the intensity of density fluctuations.  Suppose that, with a probability that we must calculate, the material in the neighborhood of this event contains a string of relatively loose molecules that provides a route along which the ``vacancy'' and the ``interstitial'' can propagate.  Our  strategy is to estimate the height of the free-energy barrier over which this system must fluctuate in order for it to become energetically favorable for the string to lengthen without bound.  When that happens, we postulate that a dynamically stable STZ has formed. In short, we propose to solve a nucleation problem where the reaction coordinate is the length of this string.  The entropy associated with different string configurations is a measure of the number of routes across this energy barrier and, therefore (see \cite{JSL69}), reduces the free energy of the barrier for purposes of computing the nucleation rate. 

To describe the string model outlined in the preceding paragraphs, we let the string have length $N$ in units of some characteristic molecular length, say, $\ell$, and suppose that it occupies a region of size $R$ in the neighborhood of the emerging STZ.  The total excess free energy of the system consists of several parts, which we denote:
\begin{equation}
\label{DeltaG1}
\Delta G(N,R,T)= \Delta G_{\infty} + N\,e_0  - T\,S(N,R)+ E_{int}(N,R).
\end{equation}
The first term, $\Delta G_{\infty}$, is the bare activation energy for the transition, that is, the energy required to form the ``vacancy'' and the ``interstitial.'' Until these two defects separate from each other, it will be energetically favorable for them simply to recombine; thus, especially at low temperatures, we need the string to enable the system to reach a stable configuration. At high enough temperatures, $\Delta G_{\infty}$ ought to become the ordinary Arrhenius activation energy.

The remaining terms on the right-hand side of Eq.(\ref{DeltaG1}) describe the string. $e_0$ is the energy per unit step along it. $S(N,R)$ is its entropy, which we obtain by computing the number of random walks of $N$ steps extending a distance $R$. In the limit of large $N$, the number of such walks, say $W(N,R)$, is approximately
\begin{equation}
\label{walks}
W(N,R) \approx {\rm constant} \times q^N\,\exp\,\left(- {R^2\over 2\,N\,\ell^2}\right), 
\end{equation}
where $q$ is the number of choices that the walk can make at each step. Thus,  
\begin{equation}
\label{Sapprox}
S(N,R)\approx \nu\,k_B\,N  - k_B\,{R^2\over 2\,N\,\ell^2};~~~~\nu = \ln q.
\end{equation}

The last term in Eq.(\ref{DeltaG1}), $E_{int}(N,R)$, is a repulsive interaction energy that accounts for the fact that no two pieces of the string can occupy the same position at the same time.  This part of the analysis resembles Flory's calculation of excluded-volume effects in polymers.\cite{FLORY}  Following Flory, we assume that $E_{int}(N,R)$ is approximately the square of the string density integrated over the volume occupied by the string.  Therefore, using Flory's mean-field approximation, also in the limit of large $N$, we write 
\begin{equation}
\label{Eint}
E_{int}(N,R) \approx k_B\,T_{int}\,{N^2\,\ell^d\over R^d},
\end{equation}
where $k_B\,T_{int}$ is a repulsive energy (which contains dimensionless geometric factors) and $d$ is the dimensionality of the space in which this string exists. As we shall argue, it is not necessarily true that $d=3$.

Combining these terms, we have
\begin{eqnarray}
\label{DeltaG2}
{\Delta G(N,R,T)\over k_B}&\approx& {\Delta G_{\infty}\over k_B} - \nu\,N\,(T-T_0)\cr &+& T\,{R^2\over 2\,N\,\ell^2}+ T_{int}\,{N^2\,\ell^d\over R^d},
\end{eqnarray}
where $T_0\equiv e_0/(\nu\,k_B)$.  The activation barrier is a saddle point in the $N$, $R$ plane. That is, it is a minimum of $\Delta G(N,R,T)$ as a function of $R$ (the smallest free energy for fixed $N$) and a maximum as a function of $N$ (the highest point along the reaction path).  The two $R$-dependent terms have a minimum at $R = R^*(N,T)$, where 
\begin{equation}
\label{Rstar}
[R^*(N,T)]^{d+2} \propto {N^3\over T},
\end{equation}
which is the Flory expression for the swelling of a $d$-dimensional polymer chain. Inserting this result into the $R$-dependent terms in (\ref{DeltaG2}), we find that the activation energy has the following form as a function of $N$:
\begin{eqnarray}
\label{DeltaG3}
\Delta G^*(N,T)&=& \Delta G(N,R^*,T)\cr&\approx& \Delta G_{\infty}+ {\rm constant} \times T^{d/(d+2)}\,N^{(4-d)/(d+2)}\cr&-& \nu\,N\,k_B\,(T-T_0).
\end{eqnarray}
The second term on the right-hand side is positive and, for $1<d<4$, is dominant for small enough $N$; the third term dominates at large $N$.  For $T >T_0$, the activation energy goes through a maximum at $N=N^*(T)$, where 
\begin{equation}
\label{Nstar}
N^*(T) \propto \left[{T^{d/(d+2)}\over (T-T_0)}\right]^{d+2\over 2\,(d-1)}.
\end{equation}
As in conventional nucleation theory, this fluctuation most probably will collapse for $N<N^*$, but will grow without bound if $N$ becomes larger than $N^*$.  Thus the activation energy $\Delta G^*(T)$ is the value of $\Delta G^*(N,T)$ at its maximum, that is,
\begin{eqnarray}
\label{Gstar}
\Delta G^*(T)&=&\Delta G(N^*,R^*,T)\cr & \approx& \Delta G_{\infty} + {\rm constant} \times {T^{d\over 2(d-1)} \over (T-T_0)^{4-d\over 2(d-1)}}.
\end{eqnarray}

For the naively expected case of $d=3$, these results are entirely unsatisfactory.  The $T$-dependent factor in the activation energy, $T^{3/4}/ (T-T_0)^{1/4}$, has too weak a divergence to be consistent with experimental data.  Moreover, the energy scale is wrong.  The implicit picture is one in which the string consists of a chain of $N$ monopolar, vacancy-like fluctuations, so that $e_0$ would be roughly equal to $\mu\,\ell^3$, where $\mu$ is the shear modulus and $\ell$ is the molecular length scale introduced previously. Such an energy would be of the order of an electron volt, and would correspond to a temperature $T_0$ in the range of $10^4\,K$ -- too large for our purposes by about two orders of magnitude. 

An apparently more plausible picture, and one which pertains specifically to the molecular structure of glassy materials, emerges from the concept of ``frustration-limited domains,'' introduced by Kivelson {\it et al.} \cite{KIVELSON95,TARJUS-KIVELSON}. Their idea is that, in a glass-forming material, the energetically preferred structure of small clusters of the constituent molecules is one that cannot tile an infinite space.  That is, the energetically favorable short-range order is ``frustrated'' because it cannot extend over long distances.  Thus a quenched glass may consist of many domains, inside of which the molecules have arranged themselves so as to have their preferred local coordinations -- or some approximation thereto; but these coordinations are violated on the interfaces between the domains.  Accordingly, we speculate that the STZ activity is localized on a network of two-dimensional interfaces that separate the domains.  In addition to giving us a rationale for choosing $d=2$ in the preceding analysis, this hypothesis allows the energy $e_0$ to be much smaller than before, because the fluctuations are occurring in regions where the molecules already are more loosely bound to each other than they are within the bodies of the domains.  

Choosing $d=2$ and restoring missing constants, we write Eqs. (\ref{Rstar}), (\ref{Nstar}), and (\ref{Gstar}) as follows: 
\begin{equation}
\label{Rstar2}
[R^*(N,T)]^2 \approx \left({2\,T_{int}\over T}\right)^{1/2}\,\ell^2\,N^{3/2};
\end{equation}
\begin{equation}
\label{Nstar2}
N^*(T)\approx {1\over 2\,\nu^2}\,{T_{int}\,T\over (T-T_0)^2};
\end{equation}
and
\begin{equation}
\label{Gstar2}
{\Delta G^*(T)\over k_B}\approx {\Delta G_{\infty}\over k_B} + {T_{int}\,T\over 2\,\nu\,(T-T_0)}.
\end{equation}
This result exhibits the well-known Vogel-Fulcher linear divergence at $T=T_0$; therefore we know from earlier analyses (e.g. see \cite{TARJUS-KIVELSON}) that it will agree with experimental data near the glass temperature.  

In the limit $T\to\infty$, however, Eq.(\ref{Gstar2}) predicts an excess, Arrhenius-like activation energy of magnitude $k_B\,T_{int}/2\,\nu$. (The situation is worse in three dimensions, where the activation energy grows like $T^{1/2}$ at high temperatures.) This physically unrealistic high-temperature behavior is a result of the fact that our large-$N$, mean-field approximations for the interaction energy and the entropy fail when $N$ becomes small.  In this limit, the string disappears and the interaction energy should vanish accordingly; but our approximation says that the ratio $N^2/R^2$ in Eq.(\ref{Eint}) goes to a constant.  Note that the failure of the large-$N$ approximation, by definition, occurs at the same temperature where the system switches from super-Arrhenius to simple Arrhenius behavior. This transition region apparently is where the system also switches from being solid-like to liquid-like.  

The theoretical analysis in \cite{Teff} implies that the Newtonian viscosity $\eta_N(T)$ can be written in the form $\eta_{\infty}\,\exp\,[\Delta G^*(T)/k_B\,T]$ up to slowly varying logarithmic corrections.  In that analysis as well as here, the Arrhenius part of $\Delta G^*(T)$, i.e. $\Delta G_{\infty}$, is the STZ formation energy; and the prefactor in the plastic strain rate, proportional to the Boltzmann factor $\exp\,(-\Delta G_{\infty}/k_B T)$, is the equilibrium density of STZ's.  The non-Arrhenius part of $\Delta G^*(T)$ in Eq.(\ref{Gstar2}) determines the nonequilibrium rate factor denoted in \cite{Teff} (up to a prefactor) by $\exp\,[-\alpha(T)]$; that is, $\alpha(T)\approx T_{int}/[2\,\nu\,(T-T_0)]$. Thus, in the spirit of Adam and Gibbs \cite{ADAM-GIBBS}, our super-Arrhenius rates are truly nonequilibrium quantities.  They describe transitions between near-equilibrium, inherent states and not, as sometimes has been assumed, an equilibrium distribution associated with the states themselves.  For example, it is assumed in \cite{SPAEPEN77,DUINE,TUINSTRA,DEHEY} that the equilibrium density of flow defects has a Vogel-Fulcher form; and the Cohen-Grest model \cite{COHEN-GREST} attributes super-Arrhenius behavior to percolation of liquid-like regions in equilibrated states. In our opinion, the nonequilibrium interpretation is the more natural of the two possibilities.  

In Fig.(1), we illustrate both the agreement near $T_0$ and the asymptotic disagreement at higher temperatures by plotting $\Delta G^*(T)$ obtained from our theory and from measurements of $\eta_N(T)$ for the bulk metallic glass ${\rm Zr_{41.2}\,Ti_{13.8}\,Cu_{12.5}\,Ni_{10}\,Be_{22.5}}$.  We obtained the experimental points by first fitting the high-T (above $T \cong 900\,K$) part of the data in \cite{MASUHR,LU} with the Arrhenius function $\eta_N \cong \eta_{\infty}\,\exp\,(\Delta G_{\infty}/k_B\,T)$, finding $\eta_{\infty} = 1.78\times 10^{-10}$ and $\Delta G_{\infty}/k_B T = 24300\,K$.  We then followed the procedure described in \cite{KTZ+96} and plotted $T\,\log\,(\eta/\eta_{\infty})$ as a function of the temperature. To compare this data with the prediction of Eq.({\ref{Gstar2}), we used $\nu = \ln 4$, $T_0 = 515\,K$ and $T_{int}= 2800\,K$. As expected, the theory fails above $T\cong 700\,K$. It is interesting to note, however, that the deformation measurements reported in \cite{LU} and discussed in \cite{FLP,Teff} were made in the low-temperature regime, where Eq.({\ref{Gstar2}) fits the data quite well. (Our theoretical comparisons to data for structural glasses such as those discussed in \cite{TARJUS-KIVELSON} or \cite{KTZ+96} are qualitatively the same as that shown here for the metallic glass.) 

\begin{figure}
\vspace{-1cm}
\begin{center}
\unitlength = 0.0005\textwidth
\begin{picture}(900,850)(0,0)
\put(0,0){\resizebox{900\unitlength}{!}{\includegraphics{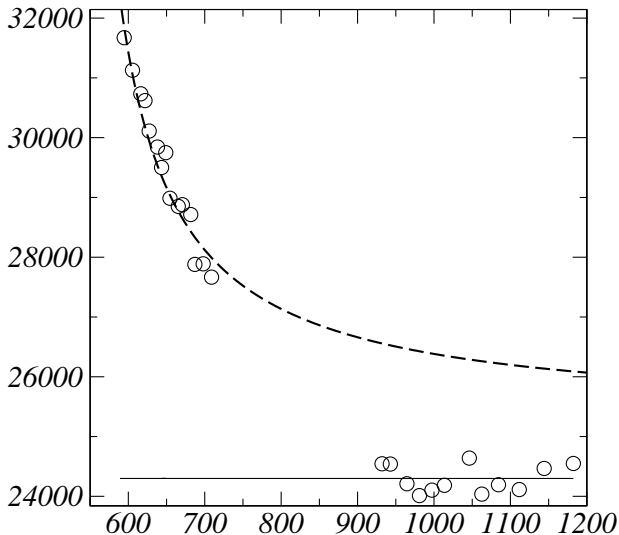}}}
\end{picture}
\label{fig:rdist}
\end{center}
\caption{Fits of metallic glass data from~\cite{MASUHR,LU}.
The dashed line is a fit of the low temperature data with our theory.
Parameters are $\Delta G_{\infty}/k_B = 24300\,K$, $T_0 = 515$, $T_{int}= 2800$, 
and $\nu = \ln 4$. 
}
\end{figure}

Although the failure of Eq.({\ref{Gstar2}) at high temperatures is clearly related to a failure of our large-$N$ approximations, we believe that simply improving the mathematics of our string model is not what is needed at this point.  Rather, it seems to us that somehow we must construct a realistic model of the transition between solid-like and liquid-like glasses using physics that so far we have not brought to bear on this problem. The picture of a solid-like glass as a three dimensional mosaic of frustration-limited domains must break down at higher temperatures, where the domains must become smaller and the active, liquid-like regions between them must occupy a larger fraction of the system.  As this happens, our strings -- or whatever replaces them -- must begin to look very different than they do in our simple two-dimensional approximation. At the most fundamental level, we believe we next must find a way to describe how a solid-like theory of the kind discussed here crosses over to a mode-coupling theory of a liquid-like glass forming material.

\begin{acknowledgments}
This research was supported primarily by U.S. Department of Energy Grant No. DE-FG03-99ER45762, and in part by the MRSEC Program of the National Science Foundation under Award No. DMR96-32716. A. Lemaitre was supported by the W. M. Keck Foundation,
and the NSF Grant No. DMR-9813752.

\end{acknowledgments}

\end{document}